\documentclass[%
 aip,
rsi,%
 amsmath,amssymb,
 reprint,%
floatfix,
]{revtex4-1}

\usepackage{graphicx}
\usepackage{dcolumn}
\usepackage{bm}
\usepackage{subcaption}
\usepackage{siunitx}
\usepackage[normalem]{ulem}
\usepackage{adjustbox}

\usepackage[usenames, dvipsnames]{color}
\usepackage[normalem]{ulem}

\begin{document}

\preprint{AIP/123-QED}

\title[X-ray Parametric Conversion Processes]{X-ray parametric down-conversion:  Challenging previous findings on the basis of improved experimental methods}

\author{C. Boemer}
 \email{christina.boemer@xfel.eu}
 \affiliation{European X-Ray Free Electron Laser, 22869 Schenefeld, Germany}
\affiliation{Department of Physics, University of Hamburg, Jungiusstrasse 9, 20355 Hamburg, Germany}
 
\author{D. Krebs}%
\affiliation{Department of Physics, University of Hamburg, Jungiusstrasse 9, 20355 Hamburg, Germany}
\affiliation{Deutsches Elektronen Synchrotron (DESY), Notkestrasse 85, 22607 Hamburg}%
\affiliation{
The Hamburg Centre for Ultrafast Imaging, Universit\"at Hamburg, 22607 Hamburg}%

\author{M. Diez} %
 \affiliation{European X-Ray Free Electron Laser, 22869 Schenefeld, Germany}
\affiliation{Department of Physics, University of Hamburg, Jungiusstrasse 9, 20355 Hamburg, Germany}

\author{N. Rohringer}%
\affiliation{Department of Physics, University of Hamburg, Jungiusstrasse 9, 20355 Hamburg, Germany}
\affiliation{Deutsches Elektronen Synchrotron (DESY), Notkestrasse 85, 22607 Hamburg}%
\affiliation{
The Hamburg Centre for Ultrafast Imaging, Universit\"at Hamburg, 22607 Hamburg}%

\author{A. Galler} %
 \affiliation{European X-Ray Free Electron Laser, 22869 Schenefeld, Germany}

\author{C. Bressler}%
 \affiliation{European X-Ray Free Electron Laser, 22869 Schenefeld, Germany}
\affiliation{ 
The Hamburg Centre for Ultrafast Imaging, Universit\"at Hamburg, 22607 Hamburg}%

\date{\today}

\begin{abstract}
X-ray parametric down-conversion is a fundamental nonlinear effect that promises a wide range of possible applications. 
Nevertheless, it has been scarcely investigated and unequivocal evidence of the effect is still missing for certain spectral regions. In particular, this is the case for down-conversion into pairs of x-ray and optical photons, which would open pathways to characterise and image valence electron-density fluctuations.
In this work, we present a systematic approach to scan the parameter space, wherein we expect to identify the effect based on its characteristic signature. 
Within the resolution of the experimental setup we cannot establish any evidence for the nonlinear effect. 
Instead, we trace the measured signals back to elastic scattering contributions, thereby challenging the interpretation of previous studies on the effect of x-ray parametric down conversion producing photons in the visible spectral range.
As a benchmark for future investigations of x-ray parametric down-conversion - both experimental and theoretical - we extract an upper bound for the effect's conversion efficiency of $10^{-11}$ within the resolution of our setup.

%
\end{abstract}

\pacs{Valid PACS appear here}
\maketitle
\begin{quotation}

\end{quotation}

\section{\label{sec:level1}Introduction:\protect\\}

Parametric processes have been extensively studied in the optical domain, where many of them are well understood and find their applications across a broad range of science and technologies \cite{photonics}.
In contrast, their extension into the x-ray domain is still a nascent field spanned by relatively few studies predominantly covering proof of principle experiments \cite{glover, McCall, tamasaku2007idler}.
An extension of parametric effects to the x-ray spectral domain has potentially high impact on the field of material science and electronic structure determination, since the effect combines imaging capabilities of coherent x-ray scattering with spectroscopy. Specifically, x-ray parametric effects that involve photons with energies in the visible spectrum, open doors to unravel microscopic details giving rise to macroscopic optical properties of materials.  

In order to push nonlinear x-ray methods towards applicability, a more detailed understanding and mapping of their parameter space is required.
In this study we investigate x-ray parametric down-conversion (XPDC) - a second order process, wherein an incident photon ('pump') is converted into a correlated pair of outgoing photons ('signal' and 'idler').
As this process is parametric, energy is conserved among the three photons (i.e., $\omega_p = \omega_s + \omega_i$) and can be distributed continuously among the generated photon pair. 
Choosing a large asymmetry ratio $\omega_s / \omega_i \gg 1$, the effect promises imaging capabilities similar to regular x-ray diffraction \cite{giacovazzo2002fundamentals} ($\omega_s \approx \omega_p$), while the coupled emission of an idler photon provides sensitivity to characteristic valence electron energies (e.g., $\hbar \omega_i \sim E_{\text{bandgap}}$). 
Ultimately, parametric conversion in the x-ray regime is thus envisioned to probe valence-electrons with atomic scale resolution \cite{tamasaku11} while holding so far unexplored potential to yield temporal information under the influence of tunable light fields. 
However, since its theoretical prediction by Freund and Levine \cite{levine70} in 1969, experimental investigations of XPDC - initiated by Eisenberger and McCall \cite{McCall} - have been challenging due to the effects' low conversion rates and prominent background contributions.
Initial experimental studies therefore focused on the simpler regime of degenerate XPDC, i.e., at \mbox{$\omega_s \approx \omega_i \approx \omega_p/2$}.
In this case, both signal and idler photons lie in the x-ray domain, a coincident detection of the photon pair is feasible. This coincident detection is a distinct proof for the simultaneous pair production and has been successfully repeated by multiple researchers \cite{yoda98, shwartz12}. 
Throughout the last 50 years further efforts have been made to observe XPDC in the non-degenerate regime - approaching high asymmetry ratios among the down-converted photon energies. Reports of idler photon energies as low as 300 eV \cite{danino1981parametric} and even 50 eV \cite{tamasaku2007idler} are on record. Notably, none of these measurements could be performed in coincidence, as the XUV idler photons are reabsorbed inside the sample. 
Similarly, recent studies \cite{shwartz17} report on x-ray frequency conversion into 2 eV optical idler photons as a proof-of-principle experiment for very specific conditions. Again, no coincidence measurement was performed even though the nonlinear material is in principle transparent at the chosen idler wavelengths \footnote{The authors attempted the coincidence detection in a separate experiment. Even though the chosen diamond sample is largely transparent in the visible, other radiative processes, in particular x-ray induced optical fluorescence from lattice vacancies and impurities, were found to be stronger by orders of magnitude, thus rendering the coincidence detection of single idler photons infeasible.}. Without this unequivocal proof by coincidence method, other characteristic XPDC signatures need to be established to provide robust evidence for the nonlinear effect.
Such a signature is given by the angular dependence of the x-ray scattering pattern resulting from the effect's specific phase-matching condition (see Sec. \ref{sec:PM}).  
In this study we employ a high resolution setup, which improves upon previous configurations \cite{shwartz17} in terms of angular resolution, and allows for a precise mapping of scattering angles and intensities. We obtain this increased resolution by use of a spatially resolving two-dimensional detector. 
Reproducing conditions the from Ref. \cite{shwartz17} we observe scattering patterns devoid of the expected XPDC signature. 
Instead, we find these scattering patterns to be exclusively composed of elastic scattering features, which notably sum up to count rates comparable to those previously reported and interpreted as the XPDC signal.
As such our results contradict the interpretation of previously observed signals as XPDC \cite{shwartz17} and suggests their reevaluation in view of elastic scattering background. 
Finally, we are able to estimate an upper bound of the nonlinear conversion efficiency on the basis of the improved setup's resolution, which yields $10^{-11}$. 

\section{Phase-matching Condition}
\label{sec:PM}
Parametric down-conversion is a nonlinear, second order process, which can conceptually be described as nonlinear diffraction \cite{levine70, McCall, yoda98, adams}: an incident pump photon is scattered and thereby converted into a correlated photon pair, namely, into a signal and an idler photon \cite{levine70, McCall}.
Frequency conversion can be observed when the phase-matching condition is fulfilled. This implies that both energy and momentum are conserved:
\begin{align}
    \omega_p &= \omega_s + \omega_i 
    \label{pm-energy}\\
    \vec k_p + \vec G &= \vec k_s + \vec k_i .
    \label{pm-momentum}
\end{align}
Here, $\omega_p$, $\omega_s$ and $\omega_i$ represent the frequencies and $\vec{k}_p$, $\vec{k}_s$ and $\vec{k}_i$ the wave-vectors for pump, signal and idler photons, respectively. $\vec G$ denotes the reciprocal lattice vector governing the diffraction process (see Figure \ref{fig:setup} b). 
\begin{figure}[h!]
    \centering
      \includegraphics[width=\linewidth]{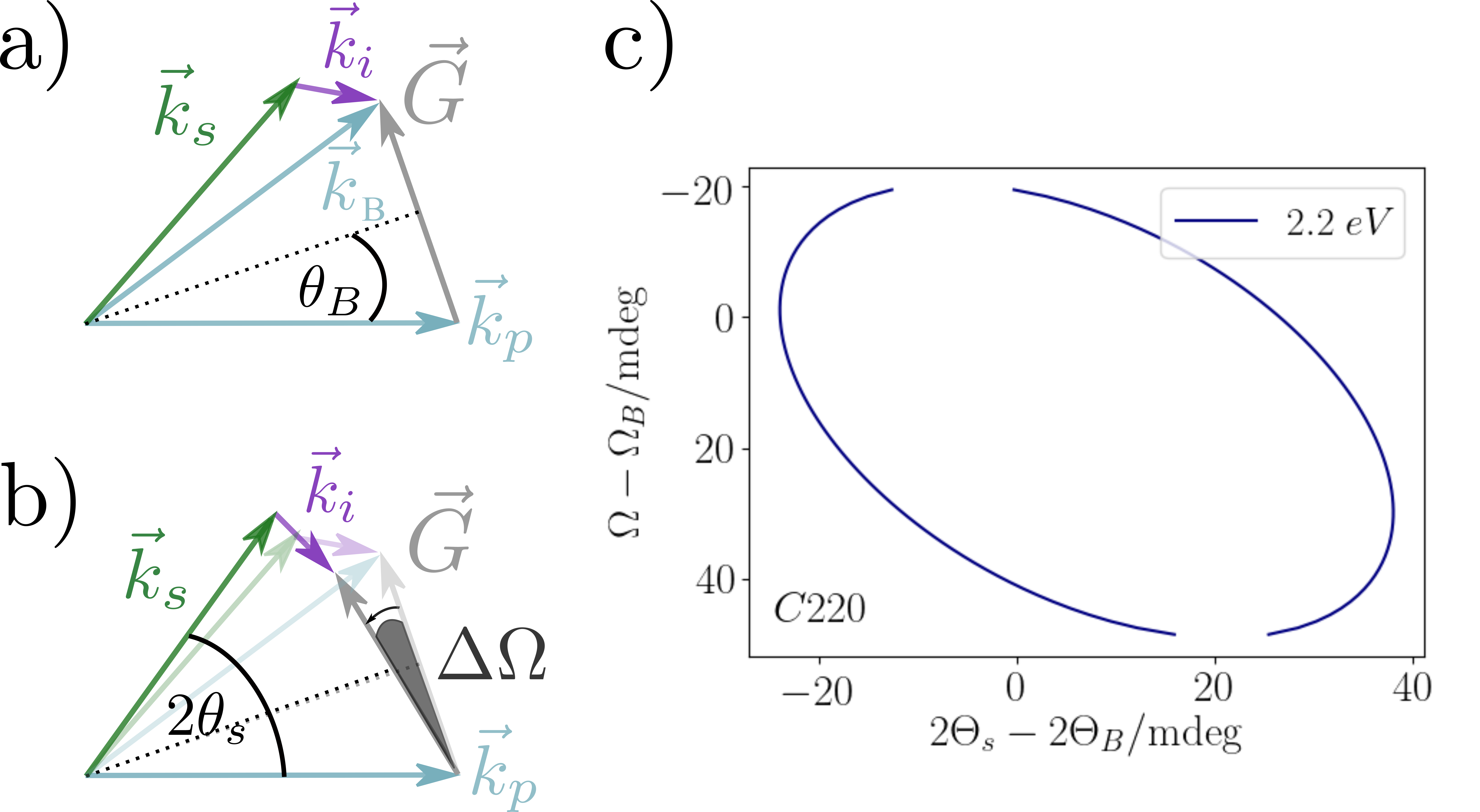}
    \caption{Basic geometry and signature of phase-matching for XPDC: the incident pump beam  with wave vector $\vec k_p$ is converted into x-ray signal ($\vec k_s$) and visible idler ($\vec k_i$) photons. 
    The process involves a reciprocal lattice vector $\vec G$ and therefore occurs close to regular Bragg conditions ($2 \theta_s \approx 2 \theta_B$) as long as $\vec k_i$ is small (a).
    By rotation of the sample ($\Delta \Omega$) the phase-matching condition is scanned (b). The resulting angular pattern is an ellipse as shown in (c) for $E_p = 11$ keV and idler energy of $E_i = 2.2$ eV.
    } 
    \label{fig:geometry}
\end{figure}
For the case of XPDC at high asymmetry ratios $\omega_s/\omega_i\gg 1$ studied in this work, the deviation of the wave vector $\vec{k}_s$ of the nonlinearly scattered x-ray photons from regular elastic scattering wave vector $\vec{k}_B$ is small. The latter yields the well known Bragg geometry ($\vec k_B = \vec k_p + \vec G$, Figure \ref{fig:geometry}(a)), whereas phase-matching including the idler photon shifts the scattering angle away from the Bragg angle $2\theta_B$.
At optical idler momentum $\frac{k_i}{k_p} \approx 10^{-4}$ this shift is of the order of few tens of mdeg (Figure \ref{fig:geometry}(c)).
Unlike Bragg scattering, parametric conversion can be achieved for a broad range of incident angles, due to the additional degrees of freedom provided by the idler photon momentum.
By rocking the sample through an angle $\Delta \Omega$, the involved momenta are reoriented - nevertheless fulfilling the phase-matching condition (Figure \ref{fig:geometry}(b)).
Tracing the x-ray scattering angle $2\theta_s$ through different sample orientations within the restrictions of Eqs.~(\ref{pm-energy}) and (\ref{pm-momentum}), defines the characteristic signature of XPDC as depicted in Fig. (\ref{fig:geometry})(b). In our experiment, we target the detection of this characteristic angular feature. 

\section{Experimental APPROACH}
\begin{figure*}[ht!]
\centering
  \includegraphics[width=0.9\linewidth, trim={0 0 0 0},clip]{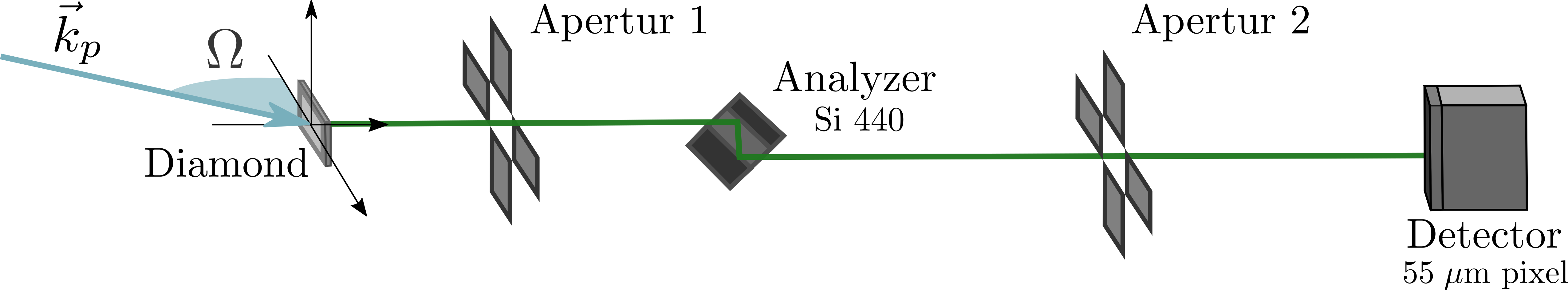}
\caption{Schematics of experimental setup: the monochromatized, collimated beam  of wave vector $\vec k_p$ diffracts (nonlinearly) from the diamond sample within the horizontal scattering plane. Behind the sample, a combination of aperture and Si 440 analyzer is used for energy discrimination. A second set of apertures reduces background before a 2D pixel detector acquires scattering patterns at an angular resolution of $2.5$ mdeg/px.}
\label{fig:setup}
\end{figure*}

To map out the parameter space of XPDC and to identify its characteristic signature, we perform rocking-curve scans, i.e. rotating the sample while detecting the scattering pattern, with a high resolution setup (Figure \ref{fig:setup}). 
The experiments are performed at beamline ID20 at ESRF (\cite{moretti2018high, huotari2017large}), where the incident beam is monochromatized by a Si(111) double crystal monochromator to a relative bandwidth of $\Delta E/E = 10^{-4}$, i.e., $1$ eV (FWHM) at the selected incident energy of $11$ keV. In this configuration, \cite{huotari2017large} the transmission of the beamline is optimal with an estimated flux of $7 \cdot 10^{13}$ ph/s.
In addition, this pump energy is far above the carbon absorption edge at $E_K = 289$ eV, avoiding resonance effects.
The beam is collimated to a degree of $1.1$ mdeg and constrained in size by horizontal and vertical apertures to $0.2 \times 0.2 \ \mathrm{mm}^2$. 
The experimental setup is a modified $\Omega - 2\theta$ diffraction setup, where the scattering proceeds in the horizontal plane. This is likewise the polarization plane of incoming and scattered radiation ($\pi$-polarization).
The sample itself is a diamond single crystal, with $<100>$ surface cut and 500 $\mu$m thickness. The scattering signal propagates through a set of apertures, which confine the beam path in horizontal and vertical dimension. A Si 220 channel cut crystal analyzer - aligned for the 440 reflection -  is used for energy discrimination of the scattered radiation.
Notably, the analyzer reflects the signal out of the original scattering plane, yielding a vertical offset of the beam path. The overall energy resolution is determined by the combination of the first aperture ($0.2$ mm) and the crystal analyzer, which is fixed to 0.3 eV (FWHM). Downstream of the analyzer, a second set of apertures reduces background scattering. Finally the energy discriminated signal is recorded by a 2D pixel detector ($55 \ \mu$m pixel size, 256 x 256 px \cite{ponchut2011maxipix}) with photon counting capabilities.
The applied experimental geometry leads to an angular resolution of $2.5$ mdeg/pixel.
The available field of view (defined by the second aperture to $95$ mdeg) enables the acquisition of a $2\theta_s$ scan in a single shot.
The alignment proceeds in the following manner: The sample crystal is placed in the diffractometer's center of rotation with the help of an optical theodolite, such that the axis of rotation resides on the sample surface.
Subsequently, the sample and detector angles are aligned for Bragg condition and optimization is done by minimizing the rocking curve of the \textless400\textgreater \ reflection - measured to have a width of $1.4$ mdeg. 
Consecutively, apertures and channel-cut analyser are aligned on the Bragg reflected beam at the fundamental energy ($\omega_p$). 
The Bragg reflex on the detector is used as reference relative to which all following scattering angles are measured. 
After calibration the sample is realigned for Laue diffraction in the 220 orientation. Adapting to the transmission geometry, the axis of rotation is shifted to the center of the sample.
In the following rocking curve scans, the sample is rotated by $\Delta \Omega$ in a range of $\pm 100$ mdeg from the Bragg angle $\Omega_B$. The resulting distribution of scattered intensities is acquired with the energy analyzer that is set to accept different down-converted x-ray energies $\hbar \omega_s$. 

\section{Results \& Discussion}
\begin{figure*}[ht!]
\centering
  \includegraphics[width=0.8\linewidth, trim={0 0 0 0},clip]{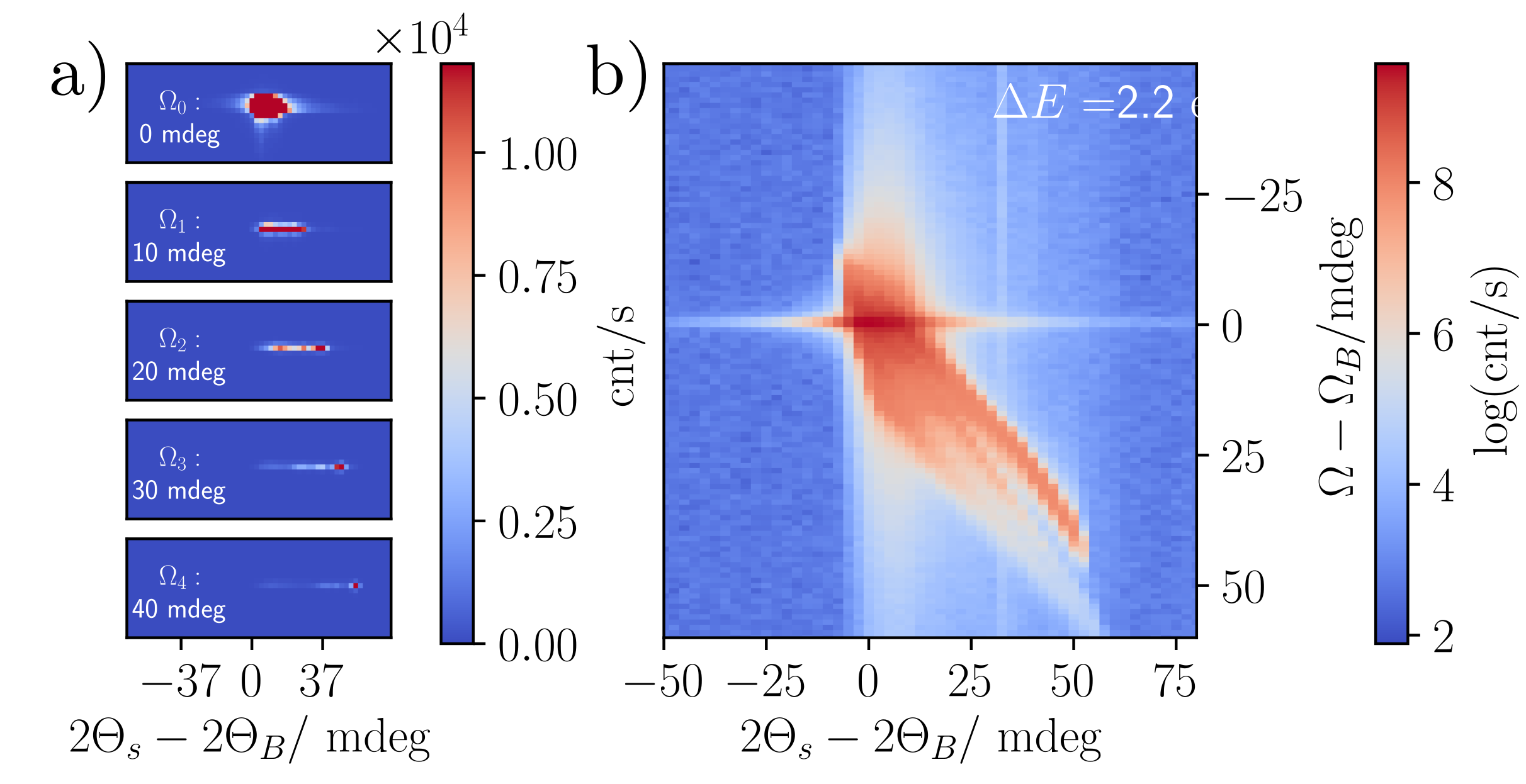}
\caption{Measured intensity distribution for rocking curve scans: the left side (a) presents detector images of scattering patterns (linear intensity scale) for various sample angles $\Delta \Omega$. Diffraction proceeds for the 220 orientation and the analyzer is detuned from $E_p$ by $\Delta E = 2.2$ eV. The right side (b) shows the condensed data set: for each sample angle $\Omega$ the detector image is integrated along the vertical dimension, which preserves the horizontal scattering information. The stack of resulting intensity distributions is shown as a map in logarithmic scale and presents the basis for further analysis.}
\label{fig:det_img_mesh}
\end{figure*}

\begin{figure}[ht!]
\centering
  \includegraphics[width=\linewidth]{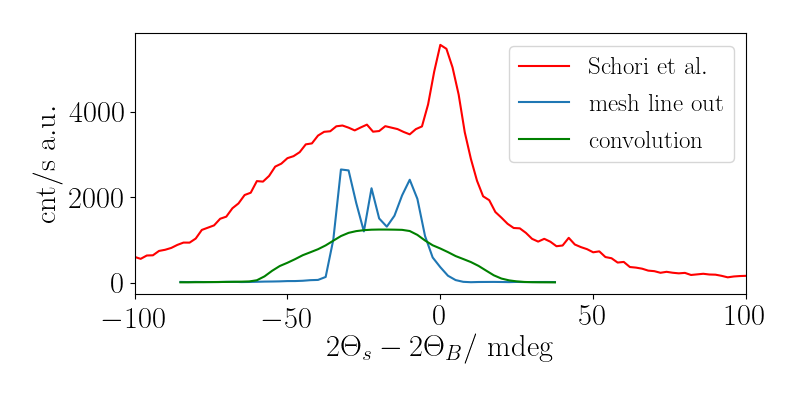}
\caption{Comparison of the data acquired in this study (blue) with the data from Ref.\ \cite{shwartz17} (red) for a same sample detuning of $\Delta \Omega = 21$ mdeg. Accounting for differences in incident flux and angular resolution, we show an accordingly convoluted version of our data (green) for clearer comparison with Ref.\ \cite{shwartz17}.}
\label{fig:comparison}
\end{figure}

\begin{figure}[h!]
\centering
  \includegraphics[width=0.8\linewidth]{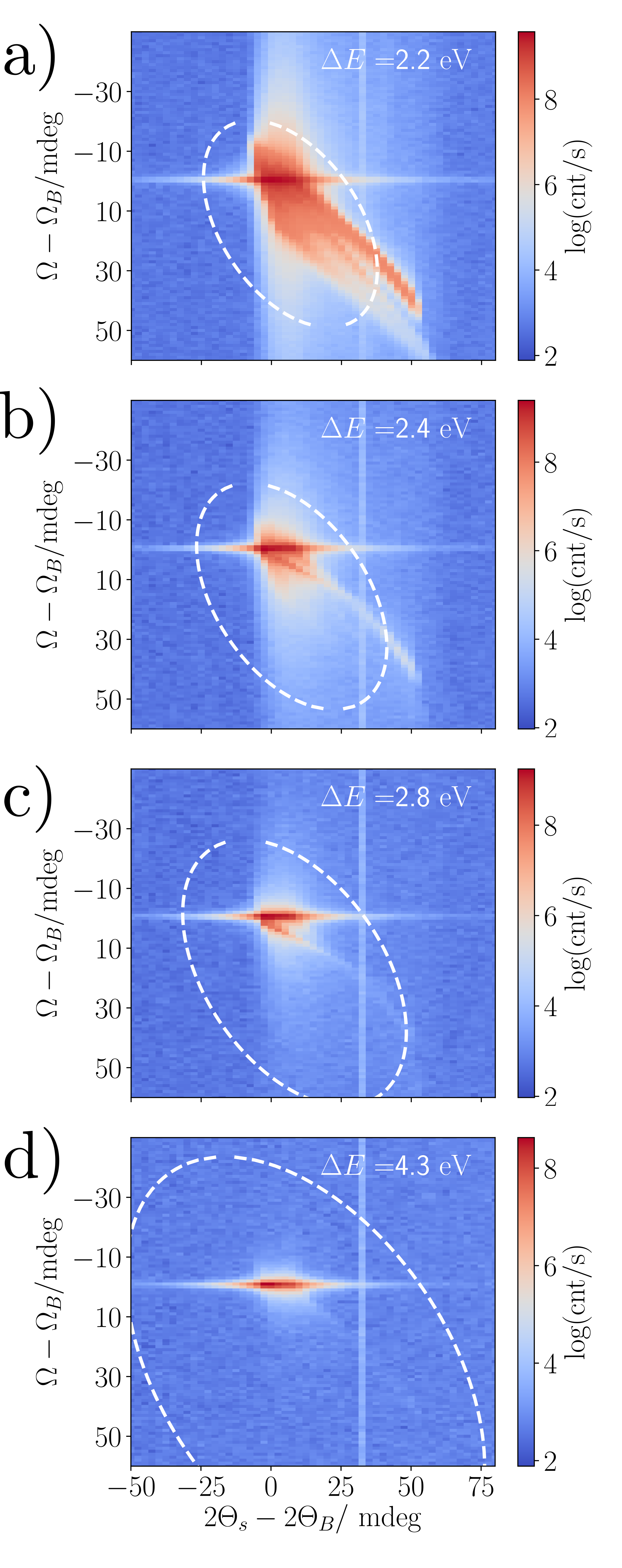}
\caption{Intensity distributions for rocking curve scans at different energy detunings $\Delta E$ for the 220 orientation. The phase-matching condition of XPDC predicts an elliptical scattering pattern (white dotted line). The data is presented on logarithmic scale. The detected intensity distribution reduces with increasing detunings from fundamental energy ($\Delta E$), remains spatially fixed and does not match the predicted scattering angles.}
\label{fig:mesh_idler}
\end{figure}

For each rocking angle $\Omega$ of the sample, the intensity distribution of the scattered radiation is recorded by the 2D detector. A series of such scattering patterns is shown on the left hand side of Figure \ref{fig:det_img_mesh} for an analyzer detuning of $\Delta E = 2.2$ eV from the incident photon energy $\omega_p$.
Starting from $\Omega = 0$ mdeg, which marks the Bragg condition for the pump energy, the series shows acquisitions for sample angles of up to $50$ mdeg rotation. 
Even though the analyzer is detuned by $2.2$ eV, the suppression of the fundamental Bragg scattering is insufficient to prevent strong overexposure of the first image.
For the following rotation angles ($\Omega$) the scattering signal shifts to higher angles in $2\theta_s$ and decreases in intensity. 
In order to obtain a more comprehensive picture, the rocking curve scan is condensed into a single data set. 
Therefore, the two-dimensional detector images are integrated along their vertical dimension and the resulting line-outs are mapped with respect to the rocking angle (Figure \ref{fig:det_img_mesh}(b)), preserving the angular resolution within the horizontal scattering plane.
The resulting rocking-curve map exhibits complex structures around the Bragg peak, which are emphasized using logarithmic scaling. 
In addition, we find a broad diffuse background ranging from $2\theta_s -2\theta_B = -10$ to $70$ mdeg, which is truncated by the influence of the second aperture.
A broken pixel at $2\theta_s - 2\theta_B = 35$ mdeg is left uncorrected as a guide to the eye. 
Based on the broad parameter range of the rocking-curve map, we can first of all set our study in context with previously reported results from Ref \cite{shwartz17}. Therein, the scattered intensity is scanned in the horizontal plane for a fixed rocking angle of the sample.
The particular sample angle ($\Delta \Omega = 21$ mdeg)  of reference \cite{shwartz17} can be extracted from Figure~\ref{fig:comparison} and is used for comparison. 
The respective line-out is shown in Figure \ref{fig:comparison} (blue) together with the reference data (red) of \cite{shwartz17}. Notably, the respective setups differ both with respect to the incident flux and the detection scheme.
In order to allow for a better comparison of the data, we plot a third line (green), which is a convolution of our line-out with a window function of 10 mdeg width.  
This accounts for the lower angular resolution achieved in Ref.\ \cite{shwartz17}, where a spatially non-resolving detector (avalanche photo diode) was framed by limiting apertures. Furthermore, we include normalization for the flux difference and find the resultant signal to be largely consistent with the reference data both in position and count rates.
This being established, we turn towards our central aim of identifying a clear signature of the parametric down-conversion process. 
Indeed, Figure \ref{fig:det_img_mesh}(b) exhibits several curve-like features, which could be considered as possible candidates of XPDC. These features need to be compared to the expected phase-matching condition, as shown in Figure \ref{fig:mesh_idler}(a) by the white dashed line.
Apparently, none of the features coincide with the predicted signature.
This suggests that the recorded signal does not result from XPDC but rather from regular elastic scattering effects. Notably, the setup and thus our mapping is sensitive towards the full distribution of the incident spectrum. Importantly, even the incoming photons with energies in the suppressed spectral tails of the applied monochromator contribute to the measured features.

This suggests that the recorded signal does not result from the nonlinear process but rather from regular elastic scattering effects.
In order to further confirm this finding, the study is extended towards rocking curve maps for higher energy detunings $\Delta E$ (Figure \ref{fig:mesh_idler}(b-d)). 
For these measurements, we again observe the discrepancy of measured features from the predicted phase matching condition. Even more striking, we find that the scattering pattern remains unchanged in angular position (in contrast to the phase-matching condition), while its intensity decreases for higher energy detuning, the latter being consistent with our interpretation in terms of elastic scattering of the spectral tail of the incoming radiation. The XPDC signal, in contrast, would not suffer the observed suppression given that the energy of the down-converted photons match the analyzer's passwidth.    
With the evidence provided by this systematic study, we conclude that the measured scattering intensity can be traced back exclusively to elastic scattering of the incident radiation.
As a matter of fact, compatible features are known from high resolution diffractometry \cite{mikhalychev2015ab,2004Pietsch-HighResDiffr}, where their origins have been identified in the effect of specific components of the experimental setup. 
Major contributions relevant to our setup stem from the spectral and angular spread of the incident radiation. 
Addressing these limitations, we further improve the setup - including an additional Si 311 high-resolution monochromator. 
\begin{figure}[ht!]
\centering
  \includegraphics[width=0.8\linewidth]{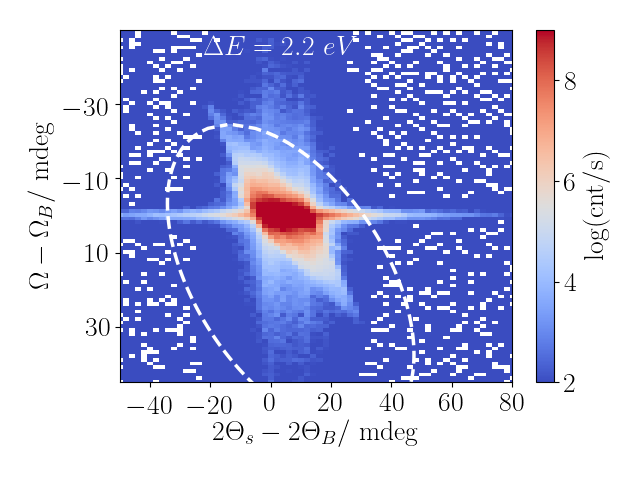}
\caption{Intensity distributions for rocking curve scan, comparable to Figure \ref{fig:mesh_idler}(a) but for modified incident beam conditions: $E_p = 10$ keV at narrower of bandwidth of $0.3$ eV. Elastic scattering is reduced, however it is still the dominant process and a clear signature for XPDC is not observed.}
\label{fig:mesh_idler_high_res}
\end{figure}
In addition to narrowing the bandwidth to $0.3$ eV FWHM, this monochromator also yields a stronger suppression of the spectral tails of the incoming undulator radiation by several orders of magnitude. With this modification, we repeated the above experiment at $\omega_p=10$ keV. We find the count rates of the central peak to remain largely unchanged while the surrounding intensity distribution is reduced, as expected. Again, the sought after signature of XPDC is not observed, thus confirming our previous finding on a more rigorous level.\\
Concluding our studies, we are able to give an order of magnitude estimate for the conversion efficiency of XPDC.
Starting with an incident flux of roughly $10^{13}$ ph/s coming from the beamline's Si 111 monochromator, the high-resolution monochromator reduces the flux by a factor of $\approx 3$.  The monochromatized beam impinges on the sample, where the XPDC conversion takes place with yet unknown efficiency. 
All down-converted radiation is collected within the bandwidth of the analyzer (0.3 eV) and the solid angle of each individual pixel (XX ??). The signal is further reduced by air absorption (reduction by factor $\approx 0.75$) and the quantum efficiency of the detector ($100 \%$ at 8 keV; $68 \%$ at 15 keV, cf. Ref.\cite{ponchut2011maxipix}).
Within this configuration, a single photon count per second per pixel would thus correspond to a conversion efficiency of $10^{-12}$.
In order to give a conservative estimate under consideration of the noise level, we apply a threshold of $10$ photons per pixel, which yields an upper bound of $10^{-11}$ for the conversion efficiency. 
Irrespective of its coarse nature, this estimate presents an important benchmark for future studies of XPDC. It indicates the minimal requirements placed on future experimental resolution, while in addition, it also serves as a point of reference for theoretical developments. In fact, there is description of x-ray optical wave-mixing under development by the authors 
\cite{krebs}, the preliminary results of which are fully compatible with our experimental estimate of the upper bound. 

\section{Conclusions}
In this study, we investigated the nonlinear effect of x-ray parametric down-conversion (XPDC) going beyond the scope of previous studies. 
More specifically, we set out to obtain an unequivocal signature of the XPDC effect, involving idler photons in the visible regime. 
To this end, we performed a systematic mapping of the parameter space, where its characteristic signature (phase-matching condition) is expected.
Despite the improved resolution of the experimental setup, no evidence for XPDC is observed. 
Instead the detected intensity distribution is shown to result solely from elastic scattering of the incoming radiation.
Based on these findings, we implemented further improvements to the setup, by decreasing the spectral bandwidth and suppressing the spectral tails of the incoming radiation. The resulting suppression of elastic effects confirms our hypothesis and provides improved contrast for the detection XPDC. The nonlinear XPDC signal is not observed. As a direct result, we are able to estimate an upper bound of the XPDC conversion efficiency amounting to $10^{-11}$ for the presented setup. \\
Overall, our findings stand in strong contradiction to the previous reportings by Schori et al.\ \cite{shwartz17}, for which our results suggest that elastic effects were interpreted as nonlinear scattering. 
More generally, simultaneously occurring elastic contributions need to be considered and suppressed as far as possible, when low efficiency nonlinear processes are investigated. 
By extension, this suggests a careful reevaluation of the findings and interpretations of previous studies\cite{shwartz17, borodin2019evidence, borodin2017high, sofer2019observation} in this field.\\
In conclusion, it remains an open challenge to measure and clearly identify x-ray parametric down-conversion into visible photons and ultimately access its potential for future applications.
\begin{acknowledgments}
This work is supported by the Deutsche Forschungsgemeinschaft (DFG), via the Collaborative Research Center SFB925 (Teilprojekt A4), and via the Cluster of Excellence 'The Hamburg Centre for Ultrafast Imaging' - EXC1074- project ID 194651731, and by European XFEL. D. Krebs gratefully acknowledges support through the Max Planck School of Photonics. Additional travel support by the Partnership for Innovation, Education and Research (PIER) for C. Boemer is gratefully acknowledged.
We are thankful to the European Synchrotron Radiation Facility (ESRF) for provision of beamtime (HC-3162, HC-3983).
Excellent support was provided by Marco Moretti, Christoph Sahle, Blanka Detlefs and Christian Henriquet at beamline ID20, and we thank Sharon Shwartz and his group members Aviad Schroi and Denis Borodin for providing us with additional beamtime and for their participation during some of the measurements. 
C. Boemer thanks Florian Otte for beamtime and scripting support. 

\end{acknowledgments}

\bibliographystyle{unsrt}
\bibliography{biblio}

\end{document}